\documentclass{PoS}
\usepackage[utf8]{inputenc}
\usepackage{amsmath}
\usepackage{axodraw}

\title{Leading logarithms for mesons and nucleons}

\ShortTitle{Leading logarithms for mesons and nucleons}

\ShortTitle{Leading logarithms for mesons and nucleons}

\author{\speaker{Johan Bijnens}
        \\
        Department of Astronomy and Theoretical Physics, Lund University,
        Sölvegatan 14A, SE 22362 Lund, Sweden\\
        E-mail: \email{bijnens@thep.lu.se}}

\author{Karol Kampf\\
        Institute of Particle and Nuclear Physics, Charles University,
        Prague, Czech Republic\\
        Email: \email{karol.kampf@mff.cuni.cz}}

\author{Alexey Vladimirov\\
        Institut f\"ur Theoretische Physik, Universit\"at Regensburg,
        D-93040 Regensburg, Germany\\
        E-mail: \email{vladimirov.aleksey@gmail.com}}

\abstract{This talks describes the work done in calculating leading logarithms
in massive effective field theories. We discuss shortly leading
logarithms in renormalizable theories and how they can be calculated using
only one-loop calculations in effective field theories. The remainder of the
talk discusses masses, decay constants, condensates and anomalous processes
in mesonic effective field theories like Chiral Perturbation Theory and
the expansion of the nucleon mass.}

\FullConference{The 8th  International Workshop on Chiral Dynamics \\
		 29 June 2015 - 03 July 2015\\
		 Pisa, Italy}

\renewcommand{\logo}
{\raisebox{-5mm}
{\parbox{\textwidth}{\begin{flushright}
LU TP 15-39\\
September 2015\\
\end{flushright}
\includegraphics{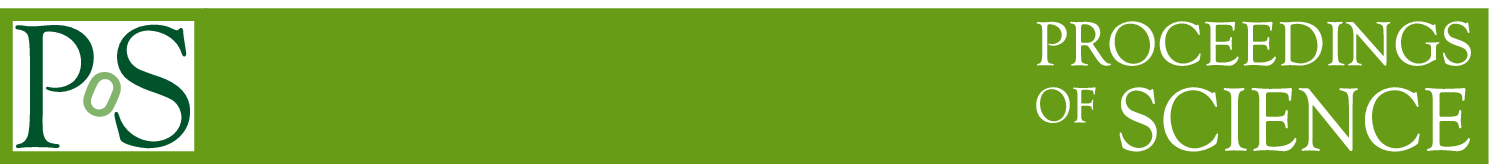}}
}}
\FullConference{{\rm To be published in the proceedings of:\\}
The 8th  International Workshop on Chiral Dynamics \\
  29 June 2015 - 03 July 2015\\
  Pisa, Italy}

\newcommand{\red}{}
\newcommand{\blue}{}
\newcommand{\dsp}{}
\newcommand{\Tr}{\mathrm{tr}}
\newcommand{\tr}{\mathrm{tr}}

\begin{document}

\section{Introduction}

The main motivation for the work described in this talk is to obtain information
on high orders in effective field theories.
The work described in this talk is published in
\cite{Bijnens:2009zi,Bijnens:2010xg,Bijnens:2012hf,Bijnens:2013yca,
Bijnens:2014ila}. There are also previous talks
containing some of these results, these include
\cite{Kampf:2013acz,Bijnens:2013zc,Vladimirov:2015msa}

The main use of our formulas is within Chiral Perturbation Theory (ChPT)
\cite{Weinberg:1978kz,Gasser:1983yg,Gasser:1984gg} but we envisage other
applications as well. A review of ChPT at loop level
is \cite{Bijnens:2006zp}, a somewhat shorter more recent review is
\cite{Bijnens:2014lea}.
The present mesonic status is reviewed in the plenary talk
by Gerhard Ecker \cite{Eckertalk}.

In Sect.~\ref{LLrenormalizable} we give a short introduction to
leading logarithms (LLs) in renormalizable field theories.
The underlying principle that allows to calculate LLs also in effective
field theories, is discussed in Sect.~\ref{LLEFT}. The remaining sections are
devoted to the different results we obtained.
Section \ref{ON} discusses a number of mesonic properties
in the massive $O(N+1)/O(N)$ model including masses, decay constants and
meson-meson scattering. The next section includes the anomaly and shows
a number of results for $\pi^0\to\gamma^*\gamma^*$ and the $\gamma\pi\pi\pi$
vertex. Sect.~\ref{SUN} discusses the LL for $N$-flavour equal mass ChPT.
Sect.~\ref{nucleon} discusses the extensions of the arguments needed for
baryon properties and the nucleon mass with results up to seven loops.
There are many more results in the papers, with similar conclusions, not
discussed here due to lack of time.

\section{Leading logarithms in renormalizable theories}
\label{LLrenormalizable}

The term leading logarithms (LLs) is used for many different things. In
this talk
it means the leading dependence on the subtraction scale $\mu$. For a
dimensionless observable
$F$ depending on a single scale $M$, Quantum field theory (QFT) tells us that
the dependence on the subtraction scale is via $L\equiv\log(\mu/M)$ in the
form
\begin{equation}
\label{Fgeneral}
F= F_0 + \left(F_1^1 {\red L} + F^1_0\right) +
\left( F_2^2{\red L^2} + F^2_1 L + F^2_0\right)
+ \left(F_3^3{\red  L^3} +\cdots\right)+\cdots
\end{equation}
Here $F^\ell_m$ means the $L^m$ contribution at order $\ell$ in
the expansion. The LLs are the terms $F^\ell_\ell L^\ell$. In QFT these terms
can be more easily calculated than the remainder. The reason is that an
observable cannot depend on the subtraction scale
$\mu(DF/d\mu)\equiv 0$ and that ultra-violet divergences in QFT are always
local. We rewrite (\ref{Fgeneral}) for the case of a renormalizable theory
with an expansion in $\alpha$
\begin{equation}
\label{Frenormalizable}
F = \alpha +\left( f_1^1 \alpha^2 L + f^1_0 \alpha^2\right) +
\left( f^2_2 \alpha^3 L^2
+f^2_1 \alpha^3 L + f^2_0 \alpha^3\right)
 +\left(f^3_3\alpha^4 L^3 + \cdots\right)+\cdots 
\end{equation}
Taking $\mu(d/d\mu)$ of (\ref{Frenormalizable}), and setting it to zero
using the beta-function
$\mu(d\alpha/d\mu)=\beta_1\alpha^2+\beta_2\alpha^3+\cdots$, gives
\begin{equation}
\left(\beta_1+f^1_1\right)\alpha^2
+\left(2\beta_1 f^1_1+2f^2_2\right)\alpha^3 L
+\left(\beta_2+2\beta_1 f^1_0+f^2_1\right)\alpha^3
+\left(3\beta_1 f^2_2+3f^3_3\right)\alpha^4 L^2
+\cdots = 0\,.
\end{equation}
The terms with highest power in $L$ at each order in $\alpha$ lead to
\begin{eqnarray}
f^1_1&=& -\beta_1,~f^2_2= \beta_1^2,~
f^3_3=-\beta_1^3,\ldots\qquad
\Longrightarrow
\nonumber\\
F(M)&=& \alpha\left(1-\alpha\beta_1 L+(\alpha\beta_1 L)^2
-(\alpha\beta_1 L)^3+\cdots\right)+\cdots
\nonumber\\
&=&\dsp{\red\frac{\alpha(\mu)}{1+\alpha(\mu)\beta_1 \log(\mu/M)}}
     +\cdots = \alpha(M)+\cdots
\nonumber
\end{eqnarray}
In the last line we showed first explicitly $\alpha=\alpha(\mu)$ and
how the LL in this case can be absorbed into a running coupling
constant. The argument can be generalized to sub-leading logarithms and extended
in general to the renormalization group.

An important part in this derivation is that the underlying theory
is the same at all orders. $\alpha$ is the same in all terms.
This reasoning is no longer true in effective field theories. 
In effective field theory we have a different Lagrangian at each order,
with different and new coupling constants.

\section{Weinberg's argument}
\label{LLEFT}

However, even if the argument used in Sect.~\ref{LLrenormalizable} no longer
holds, some possibilities remain. Weinberg \cite{Weinberg:1978kz} pointed
out that two-loop leading logarithms can be calculated using only one-loop
calculations, the method is later called ``Weinberg consistency conditions''
and relates divergences from different types of diagrams.
This method was used for $\pi\pi$-scattering at two loops
\cite{Colangelo:1995np} and the general mesonic two-loop LL
structure \cite{Bijnens:1998yu}. The extension to all orders was done
in \cite{Buchler:2003vw} and later with a diagrammatic \cite{Bijnens:2010xg}
and an operator method \cite{Bijnens:2014ila}.

Let us give the argument as presented in \cite{Bijnens:2010xg}.
We introduce $\mu$, the subtraction scale, and a parameter $\hbar$ that keeps
track of the order in the expansion. Dimensional regularization is used
throughout with $d=4-w$. The bare Lagrangian is expanded
\begin{equation}
 \mathcal{L}^\mathrm{bare} =
\sum_{n\ge0}\hbar^n \mu^{-nw}\mathcal{L}^{(n)},\quad
\mathcal{L}^{(n)} = \sum_i
{c^{(n)}_{i}}\mathcal{O}_i\,,
\quad
c^{(n)}_i = \sum_{k=0,n}\frac{c^{(n)}_{ki}}{w^k}\,.
\end{equation}
The last shows how the coefficients of the Lagrangians are expanded in the
divergences.
From QFT it follows that divergences are always local and that
only the $c^{(n)}_{0i}$ have a direct $\mu$-dependence.
The $c^{(n)}_{ki}$ $k\ge1$ only depend on $\mu$ through their dependence
on the lower order
coupling constants $c^{(m<n)}_{0i}$.
The  $\ell$-loop contribution at order $\hbar^n$ can be similarly
expanded in the divergences coming from the loop integrations
\begin{equation}
L^n_\ell = \sum_{k=0,l} \frac{1}{w^k}L^n_{k\ell}\,.
\end{equation}
The remaining parts of the argument basically use that all divergences
must cancel including the non-local ones.
At one-loop level we get a contribution
\begin{equation}
\frac{1}{w}\left(\mu^{-w} L^1_{00}(\{c\}^1_1)+L^1_{11}\right)
+\mu^{-w} L^1_{00}(\{c\}^1_0)+L^1_{10}\,.
\end{equation}
Expanding $\mu^{-w}= 1-w\log\mu+\frac{1}{2}w^2\log^2\mu+\cdots$ to get the
$\log\mu$ dependence we see that this is
$-\log\mu\, L^1_{00}(\{c\}^1_1) \equiv
\red \log\mu\, L^1_{11}$ obtainable by a one-loop calculation and by canceling
the $1/w$ term we obtain the $c^1_{1i}$.
At two loop-order we get more nontrivial results. The contribution
is
\begin{align}
&\frac{1}{w^2}\left(\mu^{-2w} L^2_{00}(\{c\}^2_2)+\mu^{-w}L^2_{11}(\{c\}^1_1)
+L^2_{22}
\right)
+\frac{1}{w}\Big(\mu^{-2w} L^2_{00}(\{c\}^2_1)+\mu^{-w}L^2_{11}(\{c\}^1_0)
\nonumber\\
&+\mu^{-w}L^2_{10}(\{c\}^1_1)
+L^2_{21}
\Big)
+\left(\mu^{-2w} L^2_{00}(\{c\}^2_0)+\mu^{-w}L^2_{10}(\{c\}^1_0)
+L^2_{20}\right)\,.
\end{align}
Canceling infinities leads to two equations
\begin{equation}
 L^2_{00}(\{c\}^2_2) + L^2_{11}(\{c\}^1_1) +L^2_{22} = 0\,,\quad
{\red 2} L^2_{00}(\{c\}^2_2) + L^2_{11}(\{c\}^1_1) = 0\,.
\end{equation}
These determine the $c^2_{2i}$ and allow to fix the LL explicit
$\mu$-dependence as
$ -\frac{1}{2}L^2_{\blue1\red1}(\{c\}^1_1)\log^2\mu$. This was
essentially Weinberg's argument in \cite{Weinberg:1978kz}.
This reasoning works to all orders, the full argument in this
form can be found in \cite{Bijnens:2010xg}.

\begin{figure}
\begin{center}
\includegraphics[width=0.98\textwidth]{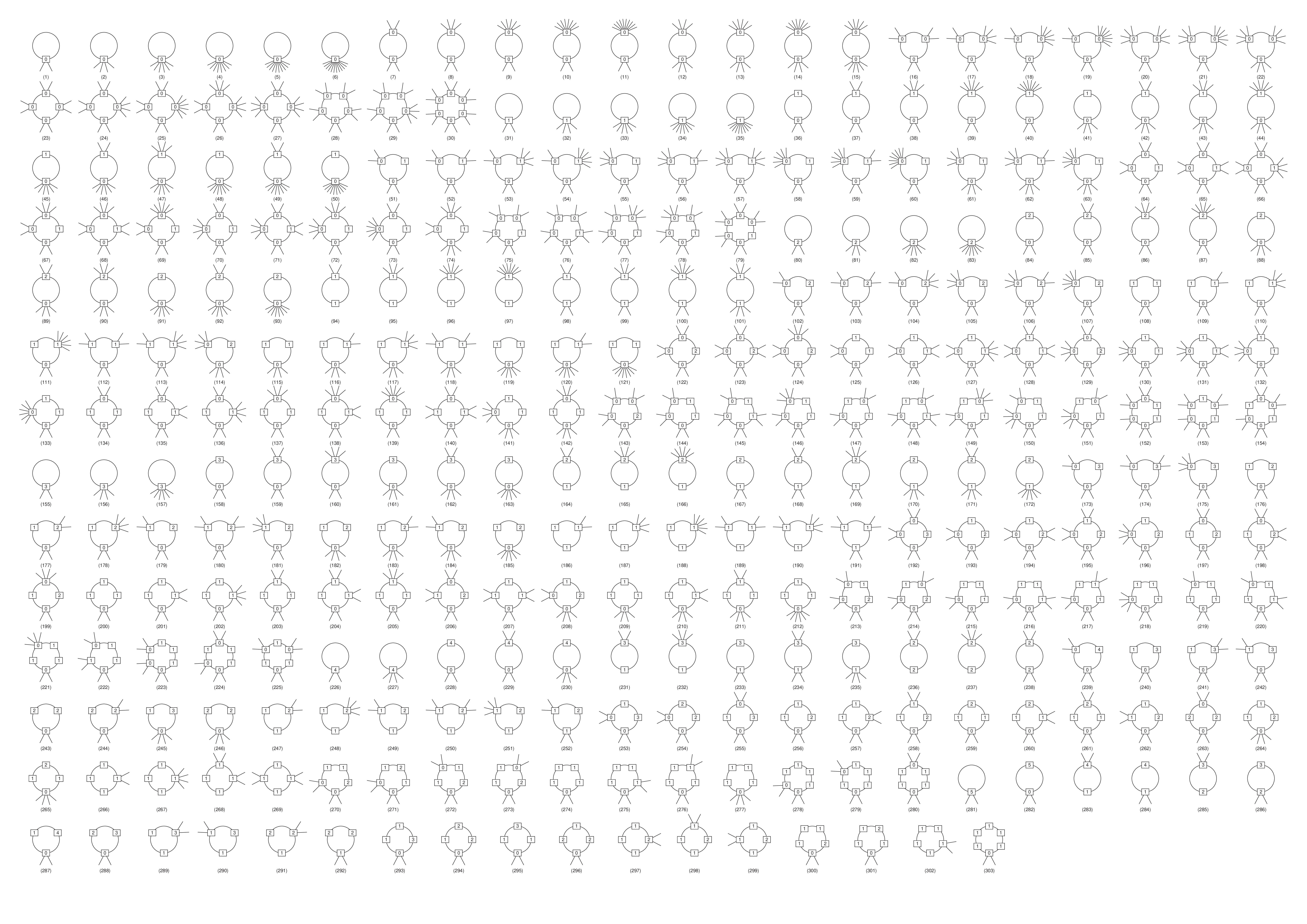}
\end{center}
\caption{The number of diagrams needed for the LL for the meson mass at
to six loops.}
\label{figLL6loop}
\end{figure}

We can thus calculate LLs using only one-loop diagrams, but for each new order
we need to take into account the new vertices. The main reason why it is
difficult to push this to higher orders is that we also get more and more
complicated diagrams at each order. This is illustrated at two-loop order
in Fig.~\ref{figLL2loop}. At higher orders the number of diagrams increases
fast. As an example we show the diagrams needed for the mass to six loops in
Fig~\ref{figLL6loop}.
\begin{figure}[tb]
\begin{center}
\begin{minipage}{0.7\textwidth}
\begin{itemize}
\item $\hbar^1$:
\setlength{\unitlength}{1pt}
\begin{picture}(50,60)(0,15)
\SetScale{1}
\SetPFont{Helvetica}{7}
\Oval(25,40)(20,20)(0)
\Line(0,20)(50,20)
\BText(25,20){0}
\end{picture}
$\Longrightarrow$
\setlength{\unitlength}{1pt}
\begin{picture}(50,60)(0,15)
\SetScale{1}
\SetPFont{Helvetica}{7}
\Line(0,20)(50,20)
\BText(25,20){1}
\end{picture}

\item $\hbar^2$:
\setlength{\unitlength}{1pt}
\begin{picture}(110,60)(0,15)
\SetScale{1}
\SetPFont{Helvetica}{7}
\Oval(25,40)(20,20)(0)
\Line(0,20)(50,20)
\BText(25,20){1}
\Oval(85,40)(20,20)(0)
\Line(60,20)(110,20)
\BText(85,20){0}
\BText(85,60){1}
\end{picture}
$\Longrightarrow$
\setlength{\unitlength}{1pt}
\begin{picture}(50,60)(0,15)
\SetScale{1}
\SetPFont{Helvetica}{7}
\Line(0,20)(50,20)
\BText(25,20){2}
\end{picture}

\item {\red but also needs} $\hbar^1$:
\setlength{\unitlength}{0.9pt}
\begin{picture}(110,60)(0,15)
\SetScale{0.9}
\SetPFont{Helvetica}{7}
\Oval(25,40)(20,20)(0)
\Line(0,10)(25,20)
\Line(0,25)(25,20)
\Line(50,10)(25,20)
\Line(50,25)(25,20)
\BText(25,20){0}
\Oval(85,40)(20,20)(0)
\Line(60,20)(110,20)
\Line(60,60)(110,60)
\BText(85,20){0}
\BText(85,60){0}
\end{picture}
$\Longrightarrow$
\setlength{\unitlength}{0.9pt}
\begin{picture}(50,60)(0,15)
\SetScale{0.9}
\SetPFont{Helvetica}{7}
\Line(0,30)(25,20)
\Line(0,10)(25,20)
\Line(50,30)(25,20)
\Line(50,10)(25,20)
\BText(25,20){1}
\end{picture}
\end{itemize}
\end{minipage}
\end{center}
\caption{\label{figLL2loop} } 
The diagrams needed for the LL up to order $\hbar^2$ for the mass.
Top line, one-loop order; Middle line: two-loop order; Bottom line: the extra
diagrams needed for the divergences of the four meson vertex at one-loop order.
This vertex is needed in the first diagram in the second line.
\begin{picture}(20,20)(0,5)\BText(10,10){n}
\end{picture} indicates a vertex from $\mathcal{L}^{(n)}$.
\end{figure}

So in practice we calculate the divergence and rewrite it in terms
of a local Lagrangian. Since we use dimensional regularization, we know that
the results have all the symmetries present in the result, so we do not need
to rewrite the Lagrangians in a nice form. This together with the fact that
we do not need to rewrite the Lagrangian in a minimal form allows to automatize
the process fully. The speed of \textsc{FORM} \cite{Vermaseren:2000nd}
played a major role
in obtaining the results described below. A small technical comment, we have
required all one-particle-irreducible diagrams to be finite, the extra
counter-terms needed for this do not affect physical results.

\section{$O(N+1)/O(N)$}
\label{ON}

In this section we discuss a few results from the massive $O(N)$ nonlinear sigma
model. The Lagrangian is given by
\begin{equation}
\mathcal{L}_{n\sigma} = \frac{F^2}{2}
\partial_\mu \Phi^T\partial^\mu \Phi+F^2 \chi^T \Phi\,. 
\end{equation}
with
$\Phi$ a real $N+1$ vector transforming as $\Phi\to O\Phi$ under $O(N+1)$
and $\Phi^T\Phi=1$.
We choose as vacuum expectation value $\langle \Phi^T \rangle =
\left(1~0 \ldots 0\right)$ and the model includes
explicit symmetry breaking via
$\chi^T = \left(M^2~0 \ldots 0\right)
$. It has thus both spontaneous
and explicit symmetry breaking with a surviving $O(N)$ global symmetry.
$N=3$ is two-flavour Chiral Perturbation Theory.
The $N$ (pseudo-)Nambu-Goldstone bosons are described by an $N$-vector $\phi$.

Calculations of this complexity need to be checked in as many ways as possible.
One good check is to use different parametrizations of $\Phi$ in terms
of $\phi$. Contributions of different diagrams can be very different with
different parametrizations, while physical quantities should be independent
of this choice. The work in \cite{Bijnens:2009zi,Bijnens:2010xg,Bijnens:2012hf}
used up to five different representations. In particular
we used the
Gasser-Leutwyler \cite{Gasser:1983yg}, Weinberg \cite{Weinberg:1968de}
and CCWZ \cite{Coleman:1969sm,Callan:1969sn}.

An example of results is the meson mass squared LLs to six loops
\cite{Bijnens:2010xg,Bijnens:2012hf}. 
\begin{equation}
\label{eqaimass}
M^2_{phys} = M^2\left(1+a_1L_M +a_2 L_M^2+\cdots\right)\,,
\quad L_M = \frac{M^2}{16\pi^2 F^2}\log\frac{ \mu^2}{\mathcal{M}^2}
\end{equation}
The usual choice for the physical scale in the logarithm is $\mathcal{M}=M$.
The coefficients are shown in Tab.~\ref{tabmass}.
\begin{table}
\begin{center}
\begin{tabular}{|c|c|l|}
\hline
i & $a_i$, $N=3$ & $a_i$ for general $N$\\
\hline
1 & $-\frac{1}{2}$        & $ 1 - \frac{N}{2}$\\[1mm]
2 & $\frac{17}{8}$       & $\frac{7}{4}
          - \frac{7 N}{4}
          + \frac{5~N^2}{8}$\\[1mm]
3 & $-\frac{103}{24}$     & $ \frac{37}{12}
          - \frac{113 N}{24}
          + \frac{15 ~N^2}{4}
          - N^3 $ \\[1mm]
4 & $\frac{24367}{1152}$ & $ \frac{839}{144}
          - \frac{1601~N}{144}
          + \frac{695~N^2}{48}
          - \frac{135~N^3}{16}
          + \frac{231~N^4}{128} $ \\[1mm]
5 & $-\frac{8821}{144}$   & $\frac{33661}{2400}
          - \frac{1151407~N}{43200}
          + \frac{197587~N^2}{4320}
          - \frac{12709~N^3}{300}
          + \frac{6271~N^4}{320}
          - \frac{7~N^5}{2} $ \\[1mm]
6 & $\frac{1922964667}{6220800}$ & $  158393809/3888000
                    - 182792131/2592000~N$\\
& & $
                    + 1046805817/7776000~N^2
                    - 17241967/103680~N^3$
\\
& & $
                    + 70046633/576000~N^4
                    - 23775/512~N^5
                    + 7293/1024~N^6$\\
\hline
\end{tabular}
\end{center}
\caption{The coefficients $a_i$ defined in
(4.2)
for the physical mass in terms of the lowest order mass.}
\label{tabmass} 
\end{table}
One question is whether one can guess at an all-order function reproducing
these coefficients. Unfortunately we did not succeed in that.
Similar results for the decay constant and vacuum expectation value
can be found in \cite{Bijnens:2009zi,Bijnens:2010xg,Bijnens:2012hf}.

\begin{figure}[tb]
\begin{center}
\includegraphics[width=0.4\textwidth]{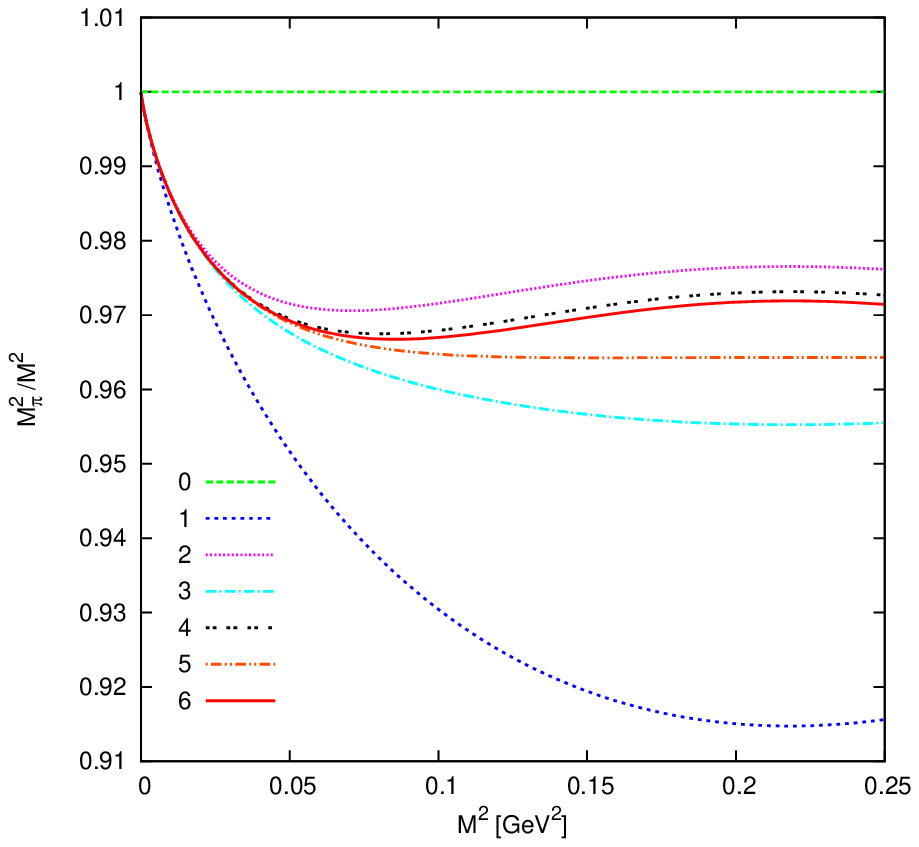}
\includegraphics[width=0.4\textwidth]{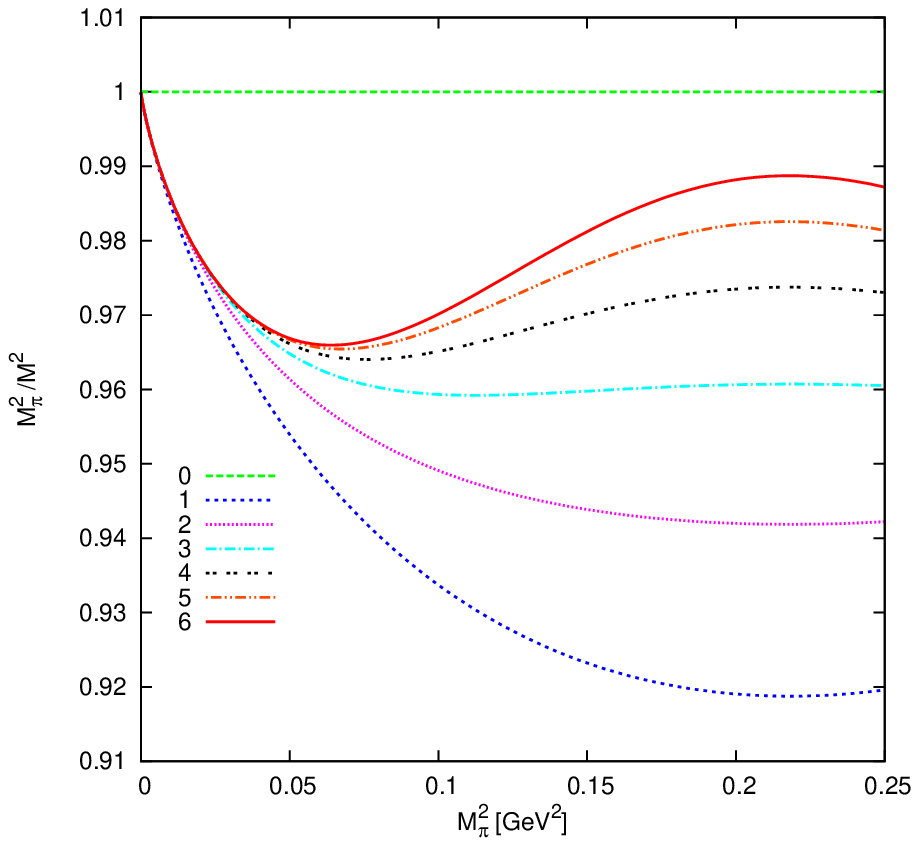}
\end{center}
\caption{
The expansion of the mass in terms of the lowest order or physical mass.
Left: $\frac{M^2_{\mathrm{phys}}}{M^2}  = 1+a_1 L_M+a_2L^2_M+a_3 L^3_M+\cdots$
with $F$ = 90 MeV, $\mu = 0.77$ GeV. Right:
$\frac{M^2_{\mathrm{phys}}}{M^2} =
1+c_1 L_{{\mathrm{phys}}}+c_2L^2_{{\mathrm{phys}}}+c_3 L^3_{{\mathrm{phys}}}+\cdots$
with
$F_\pi$ = 92 MeV, $\mu = 0.77$ GeV.} 
\label{figmassexp}
\end{figure}

\begin{figure}[tb]
\begin{center}
\includegraphics[width=0.4\textwidth]{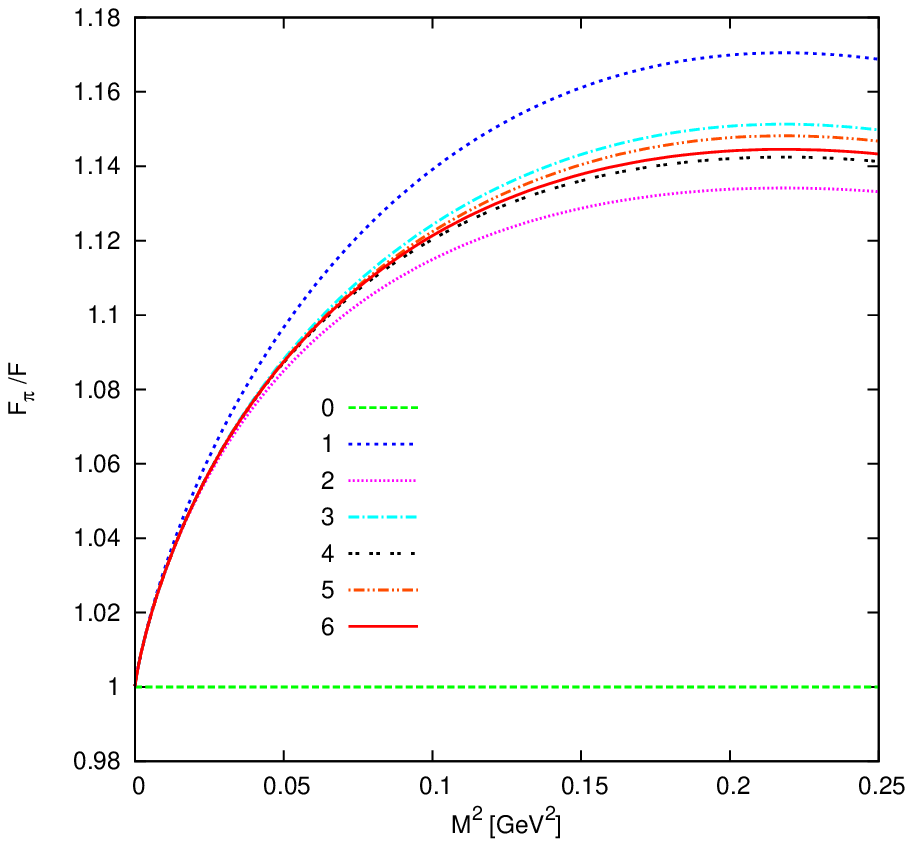}
\includegraphics[width=0.4\textwidth]{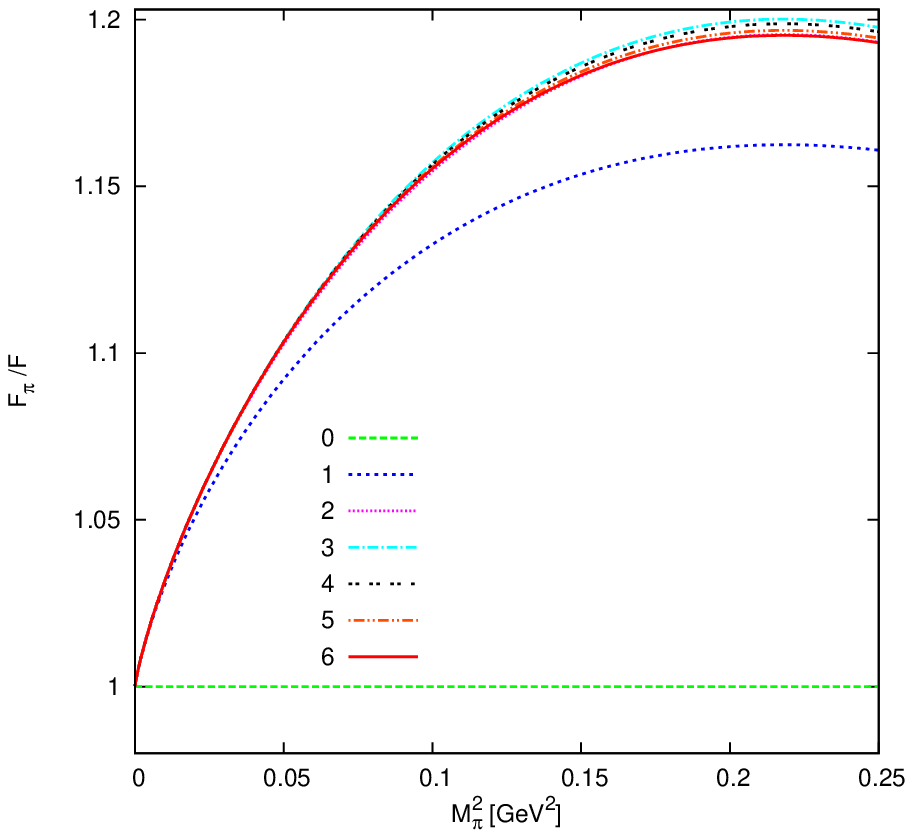}
\end{center}
\caption{
The expansion of the decay constant in terms of the lowest order or physical mass.
Left: $\frac{F_{\mathrm{phys}}}{F}  = 1+a_1 L_M+a_2L^2_M+a_3 L^3_M+\cdots$
with $F$ = 90 MeV, $\mu = 0.77$ GeV. Right:
$\frac{F_{\mathrm{phys}}}{F} =
1+c_1 L_{{\mathrm{phys}}}+c_2L^2_{{\mathrm{phys}}}+c_3 L^3_{{\mathrm{phys}}}+\cdots$
with
$F_\pi$ = 92 MeV, $\mu = 0.77$ GeV.} 
\label{figdecayexp}
\end{figure}

Using the LLs we can study how fast a series converges when rewritten in terms
of different quantities. Examples of choices are
\begin{equation}
L_M = \frac{M^2}{16\pi^2 F^2}\log\frac{\mu^2}{M^2},\quad
\tilde L_M = \frac{M_\mathrm{phys}^2}{16\pi^2 F^2}\log\frac{\mu^2}{M_\mathrm{phys}^2},\quad
L_\mathrm{phys} = \frac{M_\mathrm{phys}^2}{16\pi^2 F_\mathrm{phys}^2}\log\frac{\mu^2}{M_\mathrm{phys}^2}.
\end{equation}
For masses the expansion in $\tilde L_M$ worked best, but no general obvious
choice was found. How the choice affects the expansion is shown in
Fig.~\ref{figmassexp} when the mass is expressed in $L_M$ or $L_\mathrm{phys}$.
That the choice is not universal is illustrated by the same graphs
but for the decay constant
show in Fig.~\ref{figdecayexp}. Here the apparent quality of convergence is the
other way round.

\begin{figure}[tb]
\begin{center}
\includegraphics[width=0.4\textwidth]{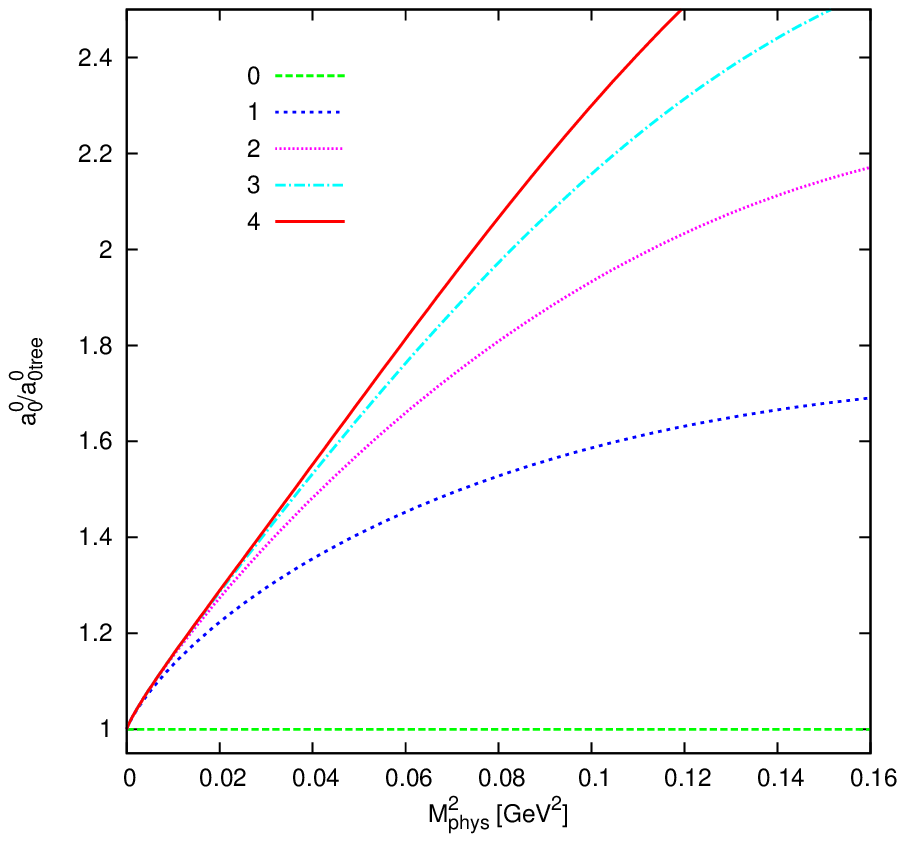}
\includegraphics[width=0.4\textwidth]{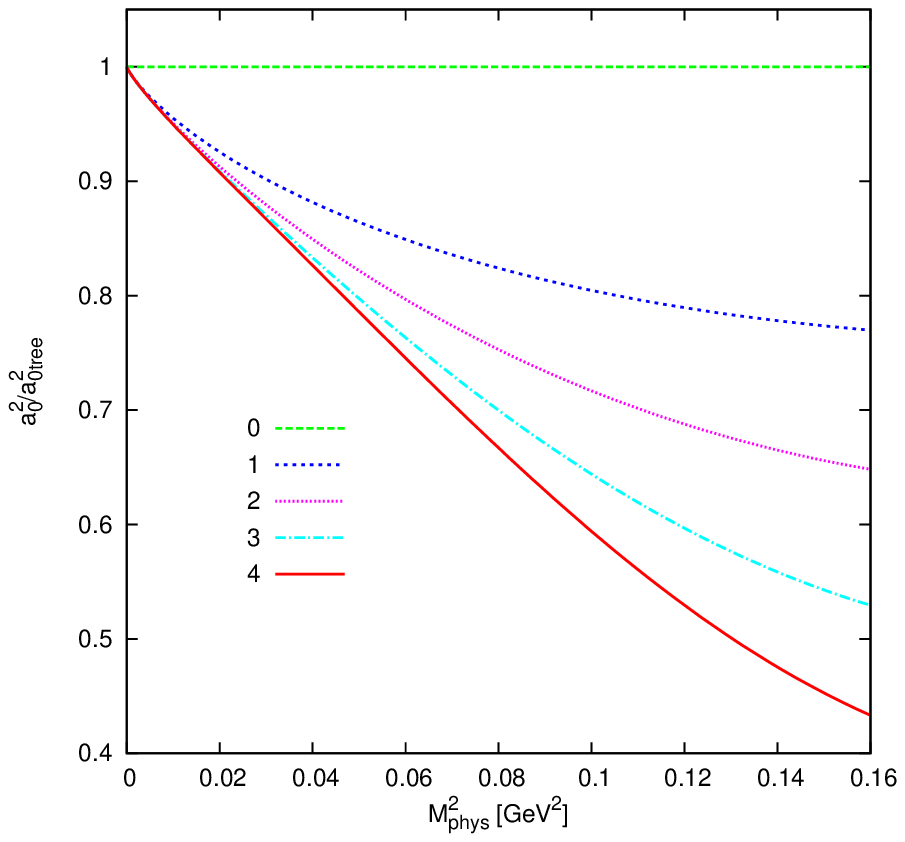}
\end{center}
\caption{
The expansion of the $a^0_0$ and $a^2_0$ $\pi\pi$-scattering lengths
with the leading logarithms.}
\label{figpipi}
\end{figure}

As a last example for this case I show the corrections to quantities which
are more dominated by the LLs. The $\pi\pi$-scattering lengths $a^0_0$
and $a^2_0$ are shown in Fig.~\ref{figpipi}.
A comparison of LL versus the full two-loop results for these is in
\cite{Bijnens:1997vq}.

More results can be found in the papers, in particular we also discussed vector
and scalar form-factors.

In the massless case tadpoles vanish and the proliferation of diagrams
does not occur in theories with four- and higher meson vertices only.
The reason is the tadpoles are responsisble for the proliferation.
This together with a parametrization of vertices using Legendre polynomials
lead to recursion relations that can be solved to very high orders
for form-factors and meson-meson scattering
\cite{Kivel:2008mf,Kivel:2009az,Koschinski:2010mr,Polyakov:2010pt}.

For the weak interactions there are also some results, for $K\to n\pi$ in
\cite{Buchler:2005jk,Buchler:2005xn}
and $K_S\to\gamma\gamma$ and
  $K_S\to\gamma l^+l^-$ \cite{Ghorbani:2014fka}. They calculated LL to two-loop
order for those processes.

\section{Anomaly for the case $O(4)/O(3)$ or $SU(2)\times SU(2)/SU(2)$}
\label{anomaly}

For anomalous processes we need to add the Wess-Zumino-Witten term to
the Lagrangian. After that we can use the same method for calculating leading
logarithms in this sector. More results and a deeper discussion can be found
in \cite{Bijnens:2012hf}. For the decay $\pi^0\to\gamma\gamma$ we find
a well known zero \cite{Donoghue:1986wv,Bijnens:1988kx}
for the LL at one-loop level but larger contributions at
higher orders. Nevertheless, the expansion converges extremely well.
Relative to lowest order the LL contributions up to
six loops are
\begin{equation}
\frac{A(\pi^0\to\gamma\gamma)_{LL}}{A(\pi^0\to\gamma\gamma)_{LO}} = 1+0-0.000372+0.000088+0.000036+0.000009+0.0000002+\ldots\,.
\end{equation}
Similarly the nonfactorizable parts with both photons off-shell is very small
and only starts at three-loop order and in the chiral limit only at four-loops.

As a last anomalous example, the amplitude for the $\gamma\pi\to\pi\pi$
vertex in terms of the usual form-factor converges as
\begin{equation}
  F_0^{3\pi LL} = (9.8-0.3+0.04+0.02+0.006+0.001+\ldots) \; \text{GeV}^{-3}
\end{equation}
We found no places with bad convergence in our work on anomalies
\cite{Bijnens:2012hf}.

\section{$SU(N)\times SU(N)/SU(N)$}
\label{SUN}

The work reported in the previous sections was mainly on the $O(N+1)/O(N)$
massive nonlinear sigma model. Other symmetry breakings are possible.
In particular $N$-flavour Chiral Perturbation Theory has the
symmetry breaking structure of $SU(N)\times SU(N)\to SU(N)$.
The methods used above can be readily generalized to this case.
In particular, the check with using different parametrizations exists as well.
We used up to four different parametrizations for a unitary matrix
and agreed with known results at two-loop orders for
masses, decay constants, vacuum expectation values \cite{Bijnens:2009qm}
and meson-meson scattering \cite{Bijnens:2011fm}.

Results for the LL for these quantities up to six loops can be found
in \cite{Bijnens:2013yca}. One of the hopes was to see if we could get a useful
leading large $N$ result since this is given by planar diagrams only.
As an example the LL for the mass coefficients are shown in
Tab.~\ref{tabmassSUN}.
\begin{table}
\begin{center}
\begin{tabular}{|c|c|c|l|}
\hline
$i$ & $a_i$ for $N=2$ & $a_i$ for $N=3$ & $a_i$ for general $N$\\
\hline
1 & $-1/2$ & $- 1/3 $        & $ - N^{-1} $\\[1mm]
2 & 17/8 & $27/8$       & $9/2\,N^{-2} - 1/2
          + 3/8\,N^2 $\\[1mm]
3 & $-103/24$ & $- 3799/648$    & $-89/3\,N^{-3}
          + 19/3\,N^{-1}
          - 37/24\,N
          - 1/12\,N^3 $ \\[1mm]
4 & 24367/1152 & $146657/2592$ & $2015/8\,N^{-4}
          - 773/12\,N^{-2}
          + 193/18
          + 121/288\,N^2$\\
& & &$
          + 41/72\,N^4$ \\[1mm]
5 & $-8821/144$ & $-\frac{27470059}{186624}$   & $- 38684/15\,N^{-5}
          + 6633/10\,N^{-3}
          - 59303/1080\,N^{-1}$\\
& & &$
          - 5077/1440\,N
          - 11327/4320\,N^3
          - 8743/34560\,N^5 $\\[1mm]
6$^*$ & $\frac{1922964667}{6220800}$ & $\frac{12902773163}{9331200}$ & $7329919/240\,N^{-6} - 1652293/240\,N^{-4}$\\
& & &$    - 4910303/15552\,N^{-2}
          + 205365409/972000$\\
& & &$
          - 69368761/7776000\,N^2
          + 14222209/2592000\,N^4$\\
& & &$
          + 3778133/3110400\,N^6
$\\
\hline
\end{tabular}
\end{center}
\caption{The coefficients of the expansion of the physical mass
in terms of $L_M$ for the $SU(N)\times SU(N)/SU(N)$ case.
Definitions as in (4.2).
}
\label{tabmassSUN}
\end{table}
Again, we did not see an all-order guess, not even for the leading $N$
coefficients.

\section{Nucleon}
\label{nucleon}

In the nucleon sector rather little has been done beyond one-loop.
Earlier work that we are aware of are the full order $p^5$ calculation
 \cite{McGovern:1998tm,McGovern:2006fm}
and order $p^6$ \cite{Schindler:2006ha,Schindler:2007dr} of the nucleon mass.
In addition there is the order $p^5$ LL correction to $g_A$,
the nucleon axial-vector coupling \cite{Bernard:2006te}.

We use as the main underlying method the heavy baryon approach,
see \cite{Bernard:1995dp} for an early review and references.
As a check we also use the relativistic formulation
\cite{Gasser:1987rb} with IR regularization \cite{Becher:1999he}.
The known problems with the latter regularization do not affect the LLs.
The results agreed. Below we only discuss the heavy baryon method.
The results of this section can be found in more detail in
\cite{Bijnens:2014ila}.

The lowest order Lagrangian is of order $p$
\begin{equation}
\mathcal{L}^{(0)}_{N\pi}=\bar N\left(i v^\mu D_\mu+g_A S^\mu u_\mu\right)N\,.
\end{equation}
As for the meson sector we need checks on our calculation. We can use
different parametrizations of the meson field as before but in addition
there are two different choices of the $p^2$ action.
The standard BKM \cite{Bernard:1995dp,Bernard:1992qa}
and the EM \cite{Ecker:1995rk} version respectively:
\begin{align}
\mathcal{L}_{\pi N}^{(1)BKM}=&\bar N_v\Big[\frac{(v\cdot D)^2-D\cdot D-ig_A \{S\cdot D,v\cdot u\}}{2M}
 +c_1\tr\left(\chi_+\right)
+\left(c_2-\frac{g_A^2}{8M}\right)(v\cdot u)^2+c_3 u\cdot u
\nonumber\\&
+\left(c_4+\frac{1}{4M}\right)i\epsilon^{\mu\nu\rho\sigma}u_\mu u_\nu v_\rho
S_\sigma \Big]N_v
\nonumber\\
\mathcal{L}^{(1)EM}_{N\pi}=& \frac{1}{M}\bar N\Big[-\frac{1}{2}
 \left(D_\mu D^\mu+i g_A\{S_\mu D^\mu,v_\nu u^\nu\}\right)+A_1 \Tr\left(u_\mu u^\mu\right)
+A_2 \Tr\left((v_\mu u^\mu)^2\right)+A_3 \Tr\left(\chi_+\right)
\nonumber\\&
+A_5i\epsilon^{\mu\nu\rho\sigma}v_\mu S_\nu u_\rho u_\sigma \Big]N\,.
\end{align}

\begin{figure}[tb]
\begin{center}
\includegraphics[width=0.7\textwidth]{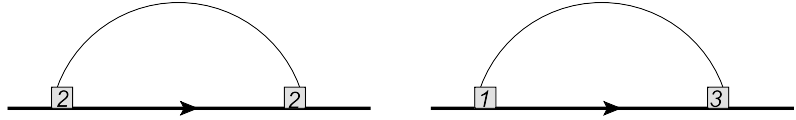}
\end{center}
\caption{Example of diagrams with different RGO but the same $p$-order.
The thick line is the nucleon, the number in the box indicate $p$-order of
the vertex. Both diagrams are order  $p^5$. The left diagram has RGO 1,
the right RGO 2}
\label{figRGO}
\end{figure}

The propagator is order $p$ but loops still add $p^2$ in the chiral counting
just as for mesons. As a consequence $p$-counting and loop counting
no longer coincide. We solve this by introducing $\hbar^2\sim p^{n+1}$
for nucleons and $\hbar^n\sim p^{n+2}$ for mesons and introduce the concept
of renormalization group order (RGO) \cite{Bijnens:2013yca}.
This is approximately the same as the maximum power of $1/w$ or $1/(d-4)$,
including the parts from the counter-terms, that can show up in a given diagram.
Diagrams at the same $p$-order can differ in RGO, an example is shown
in Fig.~\ref{figRGO}.
Note that a look at the equations 
shows that it is possible to calculate the sub-leading logarithm
if there is no contribution from tree-level diagrams at a given order.
For nucleon observables where fractional powers of the quark mass can show up
this will be the case. The example of the nucleon mass at odd $p$-orders
shows this.

We use here $M$ for the lowest order nucleon mass and $m$ for the
lowest order pion mass.
Expanding in the lowest-order logarithm $L=\frac{m^2}{(4\pi F)^2}\log\frac{\mu^2}{m^2}$ we define
\begin{align}
\label{defki}
M_{\text{phys}}=&M+k_2 \frac{m^2}{M}+k_3 \frac{\pi m^3}{(4\pi F)^2}+k_4 \frac{m^4}{(4\pi F)^2 M}\ln\frac{\mu^2}{m^2}
 +k_5 \frac{\pi
m^5}{(4\pi F)^4}\ln\frac{\mu^2}{m^2}+\cdots
\nonumber\\
=&M+\frac{m^2}{M}\sum_{n=1}^\infty k_{2n} L^{n-1}
+\pi m\frac{m^2}{(4\pi F)^2}\sum_{n=1}^\infty k_{2n+1} L^{n-1},
\end{align}
The coefficients up to $k_6$ were known from the earlier work.
We have fully calculated the coefficients up to $k_{11}$, these are shown
in Tab.~\ref{tabnucleon1}.
\begin{table}
\begin{center}
\begin{tabular}{|c||l|}
\hline $k_2$ & $-4c_1 M$
\\\hline
$k_3$ & $-\frac{3}{2}g_A^2$
\\\hline
$k_4$ & $\frac{3}{4}\left(g_A^2+(c_2+4c_3-4c_1)M\right)-3c_1M $
\\\hline
$k_5$ & $\frac{3g_A^2}{8}\left(3-16 g_A^2\right)$
\\\hline
$k_6$ & $-\frac{3}{4}\left(g_A^2+(c_2+4c_3-4c_1)M\right)+\frac{3}{2}c_1M $
\\\hline
$k_7$ & $g_A^2\left(-18 g_A^4+\frac{35 g_A^2}{4}-\frac{443}{64}\right)$
\\\hline
$k_8$ & $\frac{27}{8}\left(g_A^2+(c_2+4c_3-4 c_1)M\right)-\frac{9}{2}c_1M $
\\\hline
$k_9$ & $\frac{g_A^2}{3}\left(-116 g_A^6+\frac{2537 g_A^4}{20}-\frac{3569 g_A^2}{24}+\frac{55609}{1280}\right)$
\\ \hline
$k_{10}$ & $-\frac{257}{32}\left(g_A^2+(c_2+4c_3-4c_1)M\right)+\frac{257}{32}c_1M $
\\\hline
$k_{11}$ & $\frac{g_A^2}{2}\left(-95 g_A^8+\frac{5187407 g_A^6}{20160}-\frac{449039
   g_A^4}{945}+\frac{16733923 g_A^2}{60480}-\frac{298785521}{1935360}\right)$
\\\hline
\end{tabular}
\end{center}
\caption{The fully calculated coefficients in the LL and odd-power sub-leading
log coefficients for the nucleon mass.
The coefficients are defined in (7.3).
}
\label{tabnucleon1}
\end{table}

The Lagrangian is invariant under $g_A\leftrightarrow g_A$ and flipping the sign
of the meson field. So only even powers of $g_A$ can show up.
This is clearly visible in Tab.~\ref{tabnucleon1}.
The $k_{2n}$ have an even more peculiar structure. That only one
power of the $p^2$-Lagrangian coefficients can show up
is a consequence of the RGO counting but why only maximum $g_A^2$ shows up
is not clear to us.
If we \emph{assume} that no higher powers of $g_A$
show up we can calculate $k_{12}$ as well.
Doing that and rewriting now the logarithms in terms
of the physical pion mass leads to the simpler
coefficients $r_i$ shown in Tab.~\ref{tabnucleon2}.
\begin{table}
\begin{center}
\begin{tabular}{|c||l|}
\hline $r_2$ & $-4c_1 M$
\\\hline
$r_3$ & $-\frac{3}{2}g_A^2$
\\\hline
$r_4$ & $\frac{3}{4}\left(g_A^2+(c_2+4c_3-4c_1)M\right)-5c_1M $
\\\hline
$r_5$ & $-6 g_A^2 $
\\\hline
$r_6$ & $5c_1M $
\\\hline
$r_7$ & $\frac{g_A^2}{4}\left(-8+5g_A^2-72 g_A^4 \right)$
\\\hline
$r_8$ & $\frac{25}{3}c_1M $
\\\hline
$r_9$ & $\frac{g_A^2}{3}\left(-116 g_A^6+\frac{647 g_A^4}{20}-\frac{457 g_A^2}{12}+\frac{17}{40}\right)$
\\ \hline
$r_{10}$ & $\frac{725}{36}c_1M $
\\\hline
$r_{11}$ & $\frac{g_A^2}{2}\left(95 g_A^8-\frac{1679567 g_A^6}{20160}+\frac{451799 g_A^4}{3780}-\frac{320557
   g_A^2}{15120}+\frac{896467}{60480}\right)$
\\ \hline
$r_{12}$ & $\frac{175}{4}c_1M $
\\\hline
\end{tabular}
\end{center}
\caption{The calculated coefficients in the LL and odd-power sub-leading
log coefficients for the nucleon mass in terms of the
physical pion mass..
The coefficients are defined similar to (\protect\ref{defki}).
}
\label{tabnucleon2}
\end{table}
Taking a look at these coefficients we \emph{conjecture}
\begin{align}
\label{LLconjecture}
M=
M_\text{phys} +\frac{3}{4}m_\text{phys}^4 \frac{\log\frac{\mu^2}{m_\text{phys}^2}}{(4\pi F)^2} \left(\frac{g_A^2}{M_\text{phys}}-4c_1+c_2+4c_3\right)
 -\frac{3c_1}{{(4\pi F)^2}}\int\limits_{m^2_\text{phys}}^{\mu^2}m_\text{phys}^4(\mu')~\frac{d\mu'^2}{\mu'^2}.
\end{align}
We can now use the
known result for the pion mass discussed before
and the conjecture (\ref{LLconjecture}) to obtain  $k_{14}$ and $k_{16}$.
Using this we have calculated the LL and the odd-power sub-leading logarithms
fully to five loops and have a conjecture for the LL at 6 and 7 loops for the
nucleon mass.

\begin{figure}
\begin{center}
\includegraphics[width=0.5\textwidth]{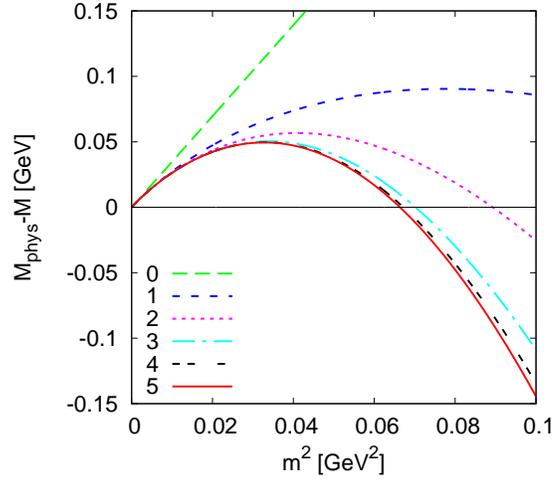}
\end{center}
\caption{
The corrections to the nucleon mass at a given loop-order from
the LL and the odd-power sub-leading logarithm.
The input parameters used are
$M = 938~\text{MeV},
 c_1 = -0.87~\text{GeV}^{-1},
 c_2 = 3.34~\text{GeV}^{-1},
 c_3 = -5.25~\text{GeV}^{-1},
 g_A= 1.25,
 \mu=0.77~\text{GeV}$ and $F = 92.4~\text{MeV}$.}
\label{fignucleonexp}
\end{figure}

\begin{figure}
\begin{center}
\includegraphics[width=0.5\textwidth]{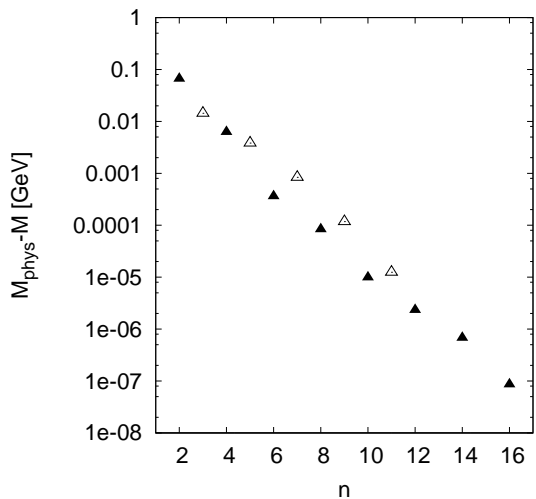}
\end{center}
\caption{The individual contribution from the $r_n$ term up $n=16$
to the nucleon mass at the physical pion mass.}
\label{fignucleonorder}
\end{figure}

Numerical results for a standard set of input parameters are shown
in Fig.\ref{fignucleonexp}.
The convergence at the physical pion mass is very good.
In Fig.~\ref{fignucleonorder} we show the contribution of all the terms we
have obtained.

\section{Conclusions}
\label{conclusions}

We discussed in this talk our recent work on leading logarithms in
massive effective field theories.
There is a rather large number of results available in the
meson sector
\cite{Bijnens:2009zi,Bijnens:2010xg,Bijnens:2012hf,Bijnens:2013yca}.
We encourage people to have a look at the (very) many tables in those
references. We welcome any all-order conjectures. 

The nucleon mass we obtained as the first result using this method in the
baryon sector \cite{Bijnens:2014ila}. 
Work is in progress for other quantities. We
had a simple conjecture for the LL at all orders in terms of the pion mass LL
but could not find a proof.

\acknowledgments
We thank our collaborators Lisa Carloni and Stefan Lanz for a joyful
collaboration and the organizers for a very pleasant and well-run conference.
This work is supported in part by the Swedish Research Council grants
621-2011-5080 and 621-2013-4287.

\bibliography{CD2015}{}
\bibliographystyle{JHEP}
%

\end{document}